\documentclass{article}

\usepackage{graphicx}
\usepackage{textcomp}

\usepackage{arxiv}

\usepackage[utf8]{inputenc} 
\usepackage[T1]{fontenc}    
\usepackage{hyperref}       
\usepackage{url}            
\usepackage{booktabs}       
\usepackage{amsfonts}       
\usepackage{nicefrac}       
\usepackage{microtype}      
\usepackage{lipsum}

\title{Problems and Solutions of Continuous Deployment: A Systematic Review}

\author{
  Antoine Proulx, Francis Raymond, Bruno Roy\\
  Département de génie informatique et génie logiciel\\
  Polytechnique Montréal\\
  Montreal, QC, Canada\\
  \texttt{\{antoine.proulx,francis.raymond,bruno-2.roy\}@polymtl.ca} \\
   \And
 Fabio Petrillo\\
  Département de Mathématique et Informatique\\
  Université du Québec à Chicoutimi\\
  Chicoutimi, QC, Canda\\
  \texttt{fabio@petrillo.com}
}

\begin{document}
\maketitle

\begin{abstract}
\textbf{Context}: The software industry needs to adapt itself to a rapidly changing market. Continuous practices (Continuous Integration, Continuous Delivery and Continuous Deployment), commonly found in Agile development processes, it is possible to deliver new features more frequently to clients, integrating of smaller features is less likely to cause conflicts than the more traditional approach of merging big features less frequently all at once. However, Continuous Deployment is no clear way on the best approaches for their implementation.\\
\textbf{Objective}: The goal of this paper is to identify the challenges and the solutions related to Continuous Deployment, and then see which of those solutions can be applied to which challenges.\\
Method: This paper is a systematic literature review of the problems and the solutions found when implementing the continuous deployment practice inside an organization. It also presents which solution can be applied to which problem. Thirty-one articles published after 2015 were analyzed for this SLR.\\
\textbf{Results}: 22 problems were grouped inside the categories \textit{Human and Organizational}, \textit{Process}, \textit{Tools}, \textit{Infrastructure}, \textit{Application Architecture} and \textit{Testing}. The 19 solutions found were grouped inside the categories \textit{Human and Organizational}, \textit{Architecture}, \textit{Process} and \textit{Tools}. Solutions have been found for 14 problems and some questions have been identified for future research.\\
\textbf{Conclusion}: this article is to serve as a reference for the practitioner who wants to find how to solve a specific challenge when implementing the continuous deployment practice. 
\end{abstract}


\section{Introduction}
\label{sec:1}
The software industry needs to adapt itself to a rapidly changing market. Traditional software development processes, like the waterfall process, are no longer appropriate. Those processes are error-prone due to the slow release rate of the product, meaning slower fixes. Because the feedback from the customers is less frequent, it is harder to evaluate if the delivered product fulfills all the needs and requirements. With Continuous practices (Continuous Integration, Continuous Delivery and Continuous Deployment), commonly found in Agile development processes, it is possible to deliver new features more frequently to clients. Another advantage of those continuous practices is that the integration of smaller features is less likely to cause conflicts than the more traditional approach of merging big features less frequently all at once. Continuous practices are rapidly growing in popularity. However, they are still quite new and there is no clear way on the best approaches for their implementation.

The goal of this paper is to identify the challenges and the solutions related to Continuous Deployment, and then see which of those solutions can be applied to which challenges. This paper should serve as a reference that identifies the solutions associated with the most common challenges that can occur when trying to implement continuous deployment.

This paper is structured as follow: Section \ref{sec:2} will go over the background, defining the concepts that we will be used in this paper and the related work, where we will go over a systematic literature mapping and two systematic literature reviews on the subject; Section \ref{sec:3} will describe the methodology used to answer our research question; Section \ref{sec:4} will present our results; Section \ref{sec:5} will be a discussion around our results, answering our research question; Section \ref{sec:6} will address threats to validity; and Section \ref{sec:7} will conclude our study.

\section{Background and Related Works}
\label{sec:2}
\subsection{Background}
In this section, we will go over what defines Continuous Integration (CI), Continuous Delivery (CDE) and Continuous Deployment (CD).

\textbf{Continuous Integration} is the process of frequently integrating the work progress of multiple developers in order to always have a version of code that is up to date. In a typical work-flow, this would translate to merging each developer's branch whenever they are functional into one. This would also include automatically building the product from the merged branch and running tests against it.

\textbf{Continuous Delivery} is a process build on top of Continuous Integration, the main code base is frequently updated, but it also has to be ready to be deployed. This means that the there needs to be automated tests running for every change in order to make sure that the software is working properly. In order to deploy a new version of their software, a team using Continuous Delivery does not necessarily need to run manual tests. There should always be a code version containing all the latest features as a valid release candidate.

\textbf{Continuous Deployment} is the process of deploying a working version of the product to the users whenever this new version successfully passes all the required tests. This process requires both Continuous Integration and Continuous Delivery with the use of a well-defined pipeline. This can lead to multiple deployments in the same day.

\subsection{Related Works}
We found three literature studies related to our research question. From those three, one is a systematic literature mapping and two are systematic literature reviews (see Table \ref{table_related_works}).

The first study related to our research question is a systematic mapping study titled ``Continuous deployment of software intensive products and services: A systematic mapping study'' \cite{Rodriguez2017}. This study aims to give an overview of Continuous Deployment based on 17 studies on the subject. The topics of analysis of this study include the general type of paper, the year of publication, the factor and definition of Continuous Deployment as well as the benefits and challenges of adopting Continuous Deployment. For instance, it was found that 72\% of the studies on the subject were of empirical nature, such as industry reports, case studies, etc. Furthermore, the study shows that 68\% of the related studies were published in the last three years, at the time of writing of this systematic mapping study, which translates to 2012-2014. This information implies that the subject of Continuous Deployment is still very new as it has started growing in popularity only in the last 5 years. The study also finds that most primary studies do not define the term Continuous Deployment. It is generally implied that Continuous Deployment defines the ability for organizations to deploy new features to the customers rapidly and at will. However, it is sometimes interchanged with the concept of Continuous Delivery, but the general idea remains the same. Most studies mention benefits such as shorter time-to-market, faster feedback and higher customer satisfaction, but no proof is provided for these benefits. It is hard to know whether or not those benefits are an actual result of Continuous Deployment or not. As for the most common challenges of implementing Continuous Deployment, they refer to the customers' unwillingness to receive continuous updates (less control over the product) as well as the increased effort required by QA teams.

The second study \cite{Laukkanen2017} is a systematic literature review (SLR) written by Eero Laukkanen, Juha Itkonen and Casper Lassenius. This paper was accepted the 11\textsc{th} of October 2016 and focused on continuous delivery while also mentioning continuous deployment. After removing the duplicates, this SLR gathered 326 papers. They then applied multiple filters and finally selected 30 articles. This SLR has three research questions that all target the continuous delivery adoption. The first question is in regards to the different \textbf{problems} that can occur during the adoption of this process. They identifies 40 different problems that can be synthesized in seven categories: ``build design, system design, integration, testing, release, human and organizational and resource problems''. Their second research question was about finding the \textbf{causes} of those problems. They gathered a total of 28 causal relationships \cite{Laukkanen2017}, the themes in which the causes seemed more frequent were in system design and testing. Their final research question was concerning the \textbf{solutions} that a team can apply in order to facilitate the adoption of a continuous delivery process. Their conclusion was that all themes have solutions but that system design and team organization (resource, human and organizational) have the most effect on a team.

This SLR is very important to this paper since their research questions are related to ours, treating challenges and solutions. However, our paper focuses on what solutions can be applied to specific challenges (relationship between a solution and a challenge). It will be interesting to see if there are any similarities between the different \textit{themes} of challenges and solutions between this study and ours.

Finally, the third study is ``Continuous Integration, Delivery and Deployment: A Systematic Review on Approaches, Tools, Challenges and Practices'', by Mojtaba Shahin, Muhammad Ali Babar, and Liming Zhu \cite{Shahin2017}. 69 papers were analyzed for this systematic literature review. Those papers were published between 2004 and June 1\textsc{st} 2016. Their authors wanted to draw a meaningful portrait of CI, CDE and CD. Because those three practices are ``highly correlated and intertwined concepts'', they did a study that encompasses all of them. They first found 30 approaches that can facilitate continuous practices, distributed in six categories: reducing build and test time; increasing visibility and awareness on build and test results in CI; supporting (semi-) automated continuous testing; detecting violations, aws and faults in CI; addressing security, scalability issues in deployment pipeline, and improve dependability and reliability of deployment process. They found that the version control systems Subversion and Git were the most used in deployment pipelines and that Jenkins was the most popular integration server. Their discussion about the tools used to implement deployment pipelines is particularly useful for our study that is centered around continuous deployment. They also identified seven critical factors related to the implementation of continuous practices, and then some practices that can help implement them. This study has a broader scope that ours, because our focus is on continuous deployment. It would be interesting to see if newer articles can provide answers to some of the questions not answered in the paper of Shanin et al., like how security is handled in deployment pipelines.

\begin{table*}[p]
  \caption{Related Works}
  \label{table_related_works}
  \resizebox{\textwidth}{!}{%
  \begin{tabular}[p]{p{6cm} p{6cm} c l}
  \hline
  Study & Authors & Numbers of included papers & Publication date \\
  \hline
  \tabularnewline
  Continuous deployment of software intensive products and services: A systematic mapping study &
  P. Rodríguez, A. Haghighatkhah, L. E. Lwakatare, S. Teppola, T. Suomalainen, J. Eskeli, T. Karvonen, P. Kuvaja, J. M. Verner, and M. Oivo &
  17 &
  January 6\textsc{th}, 2016
  \tabularnewline
  \tabularnewline
  Problems, causes and solutions when adopting continuous delivery—A systematic literature review &
  Laukkanen, J. Itkonen, and C. Lassenius &
  30 &
  October 11\textsc{th}, 2016
  \tabularnewline
  \tabularnewline
  Continuous Integration, Delivery and Deployment: A Systematic Review on Approaches, Tools, Challenges and Practices & M. Shahin, M. A. Babar, and L. Zhu &
  69 &
  March 22\textsc{nd}, 2017
  \tabularnewline
  \hline
  \end{tabular}
 }
\end{table*}

\section{Methodology}
\label{sec:3}
\begin{figure*}[!t]
\centering
\includegraphics[width=7in]{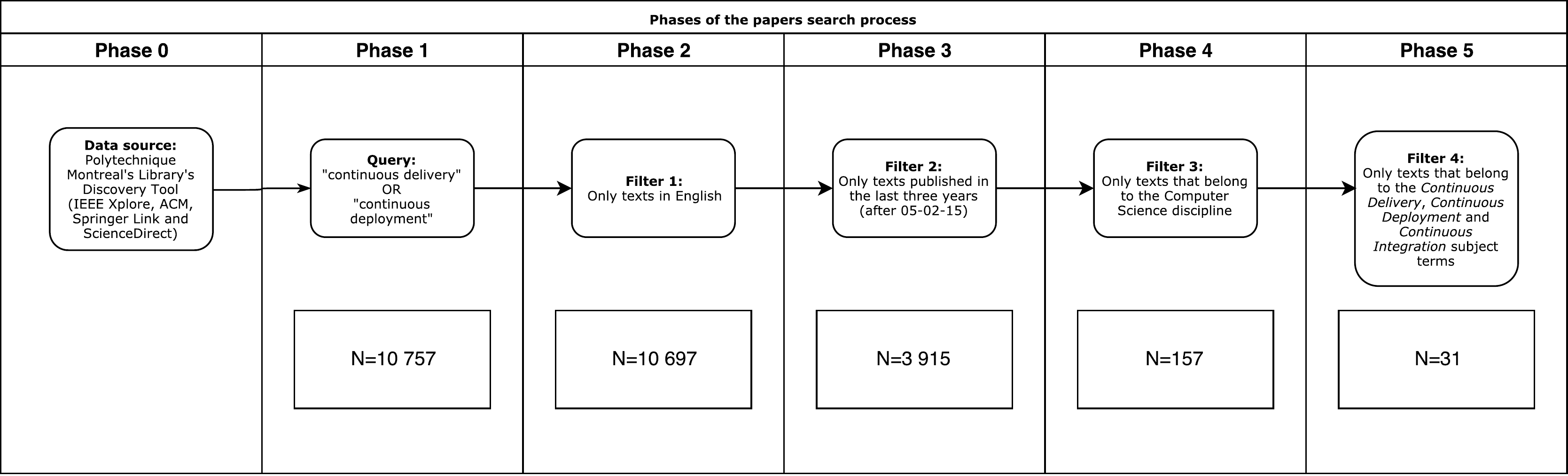}
\caption{Methodology}
\label{fig_methodology}
\end{figure*}

In order to find articles related to Continuous practices, with an emphasis on Continuous Deployment, we used Polytechnique Montreal's Library's Discovery Tool. This tool queries multiple scientific databases based on a variety of criteria, as illustrated in Figure \ref{fig_methodology}.

The query we use is the following:\\ \texttt{"continuous delivery" OR "continuous deployment"}.

We then applied multiple filters in order to reduce the number of results. The first filter we applied was to only select papers written in English. We also only wanted papers what were published after 2015 in order to get a picture that was more accurate to the current landscape. We selected the \textit{Computer Science} discipline and the \textit{Continuous Delivery}, \textit{Continuous Deployment} and \textit{Continuous Integration} subject terms.

After applying those filters, the query returned 31 papers. Those results originated from IEEE Xplore, ACM, Springer Link and ScienceDirect (see Table \ref{table_papers_found}).

\begin{table*}[p]
  \caption{Papers found}
  \label{table_papers_found}
  \resizebox{\textwidth}{!}{%
  \begin{tabular}[p]{l l}
  \hline
  Source & Paper Title \\
  
  \hline
  \tabularnewline
  IEEE Xplore & The Software Architect and DevOps \cite{Bass2018}
  \tabularnewline
  & Trends in Agile Updated: Perspectives from the Practitioners \cite{Prikladnicki2018}
  \tabularnewline
  & Built to Last or Built Too Fast? Evaluating Prediction Models for Build Times \cite{Bisong2017}
  \tabularnewline
  & Continuous Integration, Delivery and Deployment: A Systematic Review on Approaches, Tools, Challenges and Practices \cite{Shahin2017}
  \tabularnewline
  & ICSE Highlights \cite{Carver2017}
  \tabularnewline
  & The Top 10 Adages in Continuous Deployment \cite{Parnin2017}
  \tabularnewline
  & DevOps and Its Practices \cite{Zhu2016}
  \tabularnewline
  & DevOps: Making It Easy to Do the Right Thing \cite{Callanan2016}
  \tabularnewline
  & 3rd International Workshop on Release Engineering (RELENG 2015) \cite{Adams2015a}
  \tabularnewline
  & Achieving Reliable High-Frequency Releases in Cloud Environments \cite{Zhu2015}
  \tabularnewline
  & Automated Decomposition of Build Targets \cite{Vakilian2015}
  \tabularnewline
  & Continuous Delivery: Huge Benefits, but Challenges Too \cite{Chen2015a}
  \tabularnewline
  & Introducing Continuous Delivery of Mobile Apps in a Corporate Environment: A Case Study \cite{Klepper2015}
  \tabularnewline
  & Mashing Up Software Issue Management, Development, and Usage Data \cite{Mattila2015}
  \tabularnewline
  & Research Opportunities in Continuous Delivery: Reflections from Two Years' Experiences in a Large Bookmaking Company \cite{Chen2015}
  \tabularnewline
  & Securing a Deployment Pipeline \cite{Bass2015}
  \tabularnewline
  & The highways and country roads to continuous deployment \cite{Leppanen2015}
  \tabularnewline
  & The Modern Cloud-Based Platform \cite{Tilkov2015}
  \tabularnewline
  & The Practice and Future of Release Engineering: A Roundtable with Three Release Engineers \cite{Adams2015}
  \tabularnewline
  & Towards Post-Agile Development Practices through Productized Development Infrastructure \cite{Leppanen2015a}
  
  \tabularnewline
  \tabularnewline
  ACM & Bottom-up Adoption of Continuous Delivery in a Stage-Gate Managed Software Organization \cite{Laukkanen2016}
  \tabularnewline
  & Continuous Integration Beyond the Team: A Tooling Perspective on Challenges in the Automotive Industry \cite{Knauss2016}
  \tabularnewline
  & Perceived Benefits of Adopting Continuous Delivery Practices \cite{Itkonen2016}
  \tabularnewline
  & The Intersection of Continuous Deployment and Architecting Process: Practitioners' Perspectives \cite{Shahin2016}
  \tabularnewline
  & Continuous Delivery of Composite Solutions: A Case for Collaborative Software Defined PaaS Environments \cite{Austel2015}
  \tabularnewline
  & Model-based performance evaluations in continuous delivery pipelines \cite{Dlugi2015}
  \tabularnewline
  & Software Engineering Project Courses with Industrial Clients \cite{Bruegge2015}
  
  \tabularnewline
  \tabularnewline
  Springer Link & Achieving traceability in large scale continuous integration and delivery deployment, usage and validation of the eiffel framework \cite{Stahl2016}
  \tabularnewline
  & Collaborative gathering and continuous delivery of DevOps solutions through repositories \cite{Wettinger2016}
  
  \tabularnewline
  \tabularnewline
  ScienceDirect & Continuous Delivery: Overcoming adoption challenges \cite{Chen2017}
  \tabularnewline
  & Continuous deployment of software intensive products and services: A systematic mapping study \cite{Rodriguez2017}
  \tabularnewline
  & Problems, causes and solutions when adopting continuous delivery—A systematic literature review \cite{Laukkanen2017}
  
  \tabularnewline
  \hline
  \end{tabular}
}
\end{table*}

\section{Results}
\label{sec:4}
We found a total of 22 challenges and 19 solutions in regards to implementing continuous deployment. The challenges can be grouped into 6 themes that will be addressed in the following section. After going over the challenges we will cover the solutions that can be categorized into 4 themes.

\subsection{Challenges}
\begin{table*}
  \caption{Challenges}
  \label{table_challenges}
  \begin{tabular}[p]{l l l}
  \hline
  Category & Challenge & References\\
  
  \hline
  \tabularnewline
  Human and Organizational & Resistance to change & \cite{Leppanen2015}
  \tabularnewline
  & Customers don't want faster releases & \cite{Leppanen2015}
  \tabularnewline
  & Lack of confidence in the pipeline & \cite{Leppanen2015}
  \tabularnewline
  & Bigger teams make it harder to track & \cite{Chen2017}
  \tabularnewline
  & Synchronization & \cite{Parnin2017}
  
  \tabularnewline
  \tabularnewline
  Process & No emphasis on testing & \cite{Leppanen2015}, \cite{Chen2017}
  \tabularnewline
  & Split QA and Dev teams & \cite{Parnin2017}, \cite{Chen2015a}
  \tabularnewline
  & Security & \cite{Parnin2017}
  \tabularnewline
  & Different environments for production and development & \cite{Leppanen2015}
  \tabularnewline
  & Process based on non-executable models & \cite{Knauss2016}
  
  \tabularnewline
  \tabularnewline
  Tools & Tools availability & \cite{Zhu2016}, \cite{Knauss2016}, \cite{Chen2017}
  \tabularnewline
  & Tools compatibility & \cite{Parnin2017}, \cite{Zhu2016}, \cite{Chen2015a}
  \tabularnewline
  & Tools configuration & \cite{Parnin2017}, \cite{Klepper2015}
  
  \tabularnewline
  \tabularnewline
  Infrastructure & Resources to automate & \cite{Zhu2016}, \cite{Knauss2016}, \cite{Chen2017}
  \tabularnewline
  & Competition for resources & \cite{Parnin2017}, \cite{Zhu2016}, \cite{Chen2015a}
  
  \tabularnewline
  \tabularnewline
  Application Architecture & Highly coupled monolithic architecture deployment is complex & \cite{Shahin2016}
  \tabularnewline
  & Testing monolithic application & \cite{Zhu2016}, \cite{Chen2015a}
  \tabularnewline
  & Refactoring monolithic to micro-service is near impossible & \cite{Zhu2016}
  \tabularnewline
  & External resources increase chance of errors or delays & \cite{Zhu2015}
  
  \tabularnewline
  \tabularnewline
  Testing & Testing non-functional features & \cite{Leppanen2015}, \cite{Chen2017}
  \tabularnewline
  & Increase testing time & \cite{Leppanen2015}, \cite{Chen2017}
  \tabularnewline
  & Flaky tests & \cite{Laukkanen2016}
  
  \tabularnewline
  \hline
  \end{tabular}
\end{table*}

As mentioned earlier, we found a way to group the challenges into 6 categories. The categories are \textit{Human and Organizational}, \textit{Process}, \textit{Tools}, \textit{Infrastructure}, \textit{Application Architecture} and \textit{Testing}.

\subsubsection{Human and Organizational}
This section will cover the challenges related to the culture of an organization and how the organizational structure of a company can affect the implementation of CD.

The first challenge we discovered related to this category is resistance to change by the employee(s) of companies trying to implement CD. It is a well known fact that there is resistance to change whenever a company tries implementing a new approach on how to create software. This is especially true with companies trying to change to Continuous Deployment from a more traditional way of developing software, because the change can be very intimidating \cite{Leppanen2015}.

The second challenge in the category is regarding the customers. Depending on the type of software, some customers would rather get fewer releases. They would rather receive a new version of the software three to four times a year than what could be multiple versions in the same day. Part of this issue is that the customer does not trust the deployment pipeline and is scared of receiving version that are not fully tested. Also, for some industries, it's easier for the customer to have fewer releases, for example in the military or medical sector, because they cannot account for the adaptation time required for each version \cite{Leppanen2015}.

The third challenge with a human aspect is related to what we mentioned in the previous paragraph; a lack of confidence in the pipeline. The lack of trust can be observed not only from the customers, but from the developers and managers as well \cite{Leppanen2015}. Based on the research we did, the lack of confidence in the pipeline is very normal at the beginning of the implementation of Continuous Deployment. There is a lack of confidence in the pipeline (the tools used to run the automated tests and deployment), lack of confidence in the application being developed and a lack of confidence in the other members of the team \cite{Leppanen2015}.

The fourth challenge related to the organizational structure is how bigger teams have more trouble implementing Continuous Deployment. Since the release rate when using Continuous Deployment is a lot faster than the traditional approaches, it can be harder to track the work currently being done and to know exactly what is being deployed out to the customer when the development team is very large. Having smaller teams working on small subsets of the application (like with a microservice architecture) is easier to manage, especially at the start of the switch \cite{Chen2017}.

The fifth challenge that can occur when making the switch to continuous deployment is team synchronization. Since there is a new version of the application at every commit, it becomes very important to correctly order the feature implementation. For more traditional release rate (e.g. three to four times a year), it doesn't matter if you push the user interface before the logic of the application because the customers will not have the application until a formal release. This is not the case with CD; you do not want to push a user interface for a feature that is not ready yet, this would provide a very poor user experience. For larger teams, where different developers are working on parts of the same feature, organizing which part goes before which and making sure that everything is ready before pushing can be challenging \cite{Parnin2017}.

\subsubsection{Process}
This section goes over the challenges that are linked to the development process of the team that is trying to implement continuous practices. Even though continuous practices are not directly related to a specific development process, we found different aspect of the process that can make it harder to implement Continuous Deployment successfully. In some cases, depending on the current development process, it can be very hard to implement Continuous Deployment.

The first challenge occurs with teams that do not put a lot of emphasis on testing in their development process \cite{Chen2017}. It is common for teams that are not using continuous practices to lack in testing, especially automated testing \cite{Leppanen2015}. For those teams, it is harder to implement Continuous Deployment because in order to gain confidence in their pipeline they need to have good test coverage \cite{Leppanen2015} and depending on the size of the code base, this can be a monumental task.

The second issue that can arise from a more traditional process when trying to implement Continuous Deployment is in the presence of two independent teams: a designated team for testing (quality assurance) and one for developing. Continuous practices are all about speed of development, this means that you do not want a separate team that does the testing, a developer on the team should be able to complete many tasks such as writing the tests needed for a feature \cite{Parnin2017}. The challenge here is that the developers might not be familiar with the testing technologies and that the QA team is not familiar with the development methodologies. This means there is a cost associated with the training of the employees so they feel comfortable doing all the tasks needed to deploy a new version of the software \cite{Chen2015a}.

The third challenge is concerning changes that could introduce security vulnerabilities. One of the main goals of switching to Continuous Deployment is the removal of smaller independent teams (e.g. QA) and to have every developer be able to complete different tasks. It can be difficult for the average developer to know if his feature could potentially create a security vulnerability for the application. Software security can be very complex and it is hard to stay up to date with the current ways of exploiting software. This lack of knowledge and confidence can be a challenge for teams trying to implement Continuous Deployment, especially for application that deal with sensitive data \cite{Parnin2017}.

The fourth challenge in this category is the difference between the different environments. Depending on the type of software, the development environment and the production environment can be very different (e.g. embedded software for the military of medical sector). When implementing Continuous Deployment it is useful to develop on an environment that is very similar to production \cite{Leppanen2015}. If the environments are different, it increases the chances of a bug occurring and not noticing it. The configuration of the developers' environment to mimic the testing and production environments can be very complex and time consuming for an application that runs on a different platform \cite{Leppanen2015}.

The fifth challenge is when the development process is not based on executable models. Some industries, like the automotive industry, have a lot models inside their development process and it is extremely important to test them. However, those tests need to be automated to adopt CD, but sometimes, the models are described in text documents, which are not executable and cannot be automated. With executable models, there is also the need for simulators to run them \cite{Knauss2016}.

\subsubsection{Tools}
In this section we cover the challenges associated with the tools that are used when implementing Continuous Deployment.

The first challenge is the lack of tools to help set up a working deployment pipeline. Since the continuous practices are new to the industry, there aren't a lot of tools available. Most of the big companies that implement Continuous Deployment are using in house tools that were custom made for their applications \cite{Zhu2016}, \cite{Knauss2016}. This means smaller companies do not have a lot of options \cite{Chen2017}.

The second challenge in this category is finding the right tools that work well with each other. In order to fully implement a continuous deployment pipeline, you need a lot of tools (automated build system, automated test system, deployment system and more) \cite{Parnin2017}. The challenging part is find and combining tools that are compatible with each other in order to create a functional pipeline \cite{Chen2015a} \cite{Zhu2016}.

The third challenge concerning the tools is their configuration. As we mentioned earlier, since Continuous Deployment is pretty new there aren't that many tools available. Teams need to find a variety of tools in order to create a working pipeline. A challenge that arises from this combination of tools is their configuration. Most project will need their custom configuration to satisfy the business needs and most tools have their own ways of configuration that is not standardized yet. This leads to a very long and tedious process of customizing and configuring every segment of the pipeline \cite{Parnin2017}. For some projects, this amount of effort towards setting up the pipeline is one of the main reason they do not use Continuous Deployment \cite{Klepper2015}.

\subsubsection{Infrastructure}
In this section we will go over the difficulties related to the infrastructure needed to implement continuous practices.

The main challenge in this category is the computer power (i.e. resources) needed to run a fully functional pipeline. In order to implement Continuous Deployment you need to have build servers that can run the build system, run the automated test suite and finally deploy the new software. The resources requirement can be a blocker for a lot of smaller companies due to their cost \cite{Chen2017}. Also, when you have a lot of people that work on the same software, you want parallelize builds which are even more resource hungry. Some organizations have multiple branches of development for the same application and in this case, each branch needs its own pipeline, which is also resource-heavy and can prevent the adoption of Continuous Deployment practice \cite{Laukkanen2016}.

Another challenge that can arise from the lack of infrastructure is a competition between the teams in order to have the resources. Every team has their own interest and if there is not enough computer power to supply the whole company, then the teams need to share and this can lead to competition which can be unhealthy \cite{Chen2015a}.

\subsubsection{Application Architecture}
In this section we go over the challenges that are related to the architecture of the application. Some types of architecture can make it very tedious to implement CD.

The first challenge related to the application architecture is the highly coupled monolithic architecture \cite{Shahin2016}. This is a challenge because each component of the application has multiple dependencies and those dependencies need to be deployed when modifications are done on the component. It is then difficult to make changes to a component independently of the rest of the application. The application can be considered monolithic because of its code, but also because of its database \cite{Shahin2016}. The code and database both create complexity in the deployment.

The second challenge related to the application architecture is that testing an application with a monolithic architecture is harder than with a microservice architecture \cite{Chen2015a}. Most legacy system are using some variant of a monolithic architecture, and that type of architecture is not the best for Continuous Deployment because it is hard to know a change is going to have side effects on another part of the application \cite{Zhu2016}. It is hard to parallelize the development of this kind of architecture since everything is coupled together and it makes it much harder to work of a feature without impacting the work of others \cite{Zhu2016}.

The third challenge in this category is related to the one previously mentioned. It implies that refactoring a legacy system that is using a monolithic architecture into a microservice architecture can be very difficult, near impossible. Doing large refactoring on large code bases is very hard; one has to make sure that all the functionalities are still working properly. Modifying the architecture is even harder because a lot of new code has to be written. The refactoring process becomes even harder when the application does not have a large test suite, because every feature needs to be tested manually to make sure no regression happened \cite{Zhu2016}.

Lastly, it is very common for projects to use external resources. This approach makes development easier as the developers don't have to rewrite something that already exists. The downside to this is that the number of external resources greatly increases the chance of errors or delays when building, testing and deploying the project, as resource availability and stability cannot be guaranteed at all times \cite{Zhu2015}.

\subsubsection{Testing}
In this section we will cover the challenges related to testing in order to implement Continuous Deployment. A fully functional deployment pipeline is required to implement Continuous Delivery. A core part of this pipeline involves the automated tests. The tests suite run every time there is a change in the code base, to validate that the change can go out to production.

The first challenge regarding testing is related to non-functional requirements that need to be validated before production. There are many types of applications that need more testing than only functional tests. For instance, unit tests are not sufficient for video game development; manual testing is usually required in order to make sure that the game is working properly \cite{Leppanen2015}. Other types of application might require performance tests that are also harder to automate \cite{Chen2017}

The second challenge related to testing is the increase in testing time. As mentioned earlier, in order to trust the release pipeline, an application must have a thorough test suite. Running those tests on every code change can be very time consuming for larger applications and teams. This can slow down the momentum of the team and is not ideal for faster releases \cite{Chen2017}. The size of the code base is also a factor that can affect the time needed to test the application \cite{Leppanen2015}.

The third challenge is flaky tests. \emph{Flaky tests} describe tests that fail randomly. It is a challenge with Continuous Deployment practice, because the developers cannot be confident that their tests assess the correct working of the system \cite{Laukkanen2016}. Flaky tests are generally caused by an instable architecture.

\subsection{Solutions}
\begin{table*}
  \caption{Solutions}
  \label{table_solutions}
  \begin{tabular}[p]{l l l}
  \hline
  Category & Solutions & References\\
  
  \hline
  \tabularnewline
  Human and Organizational & Selling Continuous Deployment as a pain killer, a good investment & \cite{Chen2017}
  \tabularnewline
  & Assign an expert in Continuous Deployment to teams & \cite{Chen2017}
  \tabularnewline
  & Use a more collaborative approach between OEM and its suppliers & \cite{Knauss2016}
  
  \tabularnewline
  \tabularnewline
  Architecture & Micro-service architecture & \cite{Zhu2016}, \cite{Shahin2016}, \cite{Chen2017}
  \tabularnewline
  & Make micro-services backwards compatible & \cite{Callanan2016}
  \tabularnewline
  & Make database as continuously deployable unit & \cite{Shahin2016}
  \tabularnewline
  & Do not focus too much on reusability & \cite{Shahin2016}
  \tabularnewline
  & Add logging and log aggregation & \cite{Shahin2016}
  \tabularnewline
  & Isolate changes & \cite{Shahin2016}
  \tabularnewline
  & Add testability inside the architecture & \cite{Shahin2016}
  
  \tabularnewline
  \tabularnewline
  Process & Create a team that will implement Continuous Deployment & \cite{Chen2017}
  \tabularnewline
  & Continuous Deployment the Continuous Deployment implementation & \cite{Zhu2016}, \cite{Chen2017}
  \tabularnewline
  & Start Continuous Deployment implementation with easier components & \cite{Chen2017}
  \tabularnewline
  & Deploy features that aren't user facing & \cite{Parnin2017}
  \tabularnewline
  & Create a separate process to oversight security issues & \cite{Parnin2017}
  \tabularnewline
  & Use model-driven development & \cite{Knauss2016}
  
  \tabularnewline
  \tabularnewline
  Tools & Have a visual representation of the pipeline & \cite{Chen2017}
  \tabularnewline
  & Configurations should be implemented as code & \cite{Parnin2017}
  \tabularnewline
  & Availability of a turn-key solution & \cite{Klepper2015}
  
  \tabularnewline
  \hline
  \end{tabular}
\end{table*}

We will now to tackle the second part of our research question which is the solutions that a team can use when trying to implement continuous deployment. During our research we found a total of 19 solutions that we categorized into 4 themes. The themes are \textit{Human and Organizational}, \textit{Architecture}, \textit{Process} and \textit{Tools}.

\subsubsection{Human and Organizational}
In this section we will cover the solutions associated with the Human and organizational theme.

The first solution in this category is to sell continuous deployment as a pain killer, as a good investment. Implementing continuous deployment can be very challenging and very costly both in terms of money and time. The initial cost to implement this process is high. This means that it can be hard to sell the idea of switching to a continuous practice to every stakeholder, because you need to convince the development team, but also the managers and sometimes the customers. The strategy to alleviate this issue is to find for every stakeholder what they find painful in the current development process and explain how using Continuous Deployment will reduce it \cite{Chen2017}. It is usually relatively easy to convince the development team of the benefits of using Continuous Deployment, but it can be harder to convince management \cite{Chen2017}. A big concern for managers is time, because there is a big time investment in setting up the whole deployment pipeline. Once the pipeline is correctly set up, time is saved after each commit and that is what needs to be pushed to help convince \cite{Chen2017}.

The second solution in this category is to assign an expert in migrating to Continuous Deployment to teams that are currently making the switch. The goal behind this solution is that a team trying to move their applications towards Continuous Deployment should not be left alone trying to figure things out on their own \cite{Chen2017}. The expert that is assigned to the team will gain insight on how the team operates and the different applications that needs to be migrated. The expert can give advice on what the pipeline should look like. The other advantage of having an expert in the team is that it will be less likely that the team forgets a stage in the pipeline during the implementation \cite{Chen2017}. Selecting an expert can also become somewhat challenging; he must possess a deep understanding of the continuous deployment practices \cite{Chen2017}.

The third solution in this category is to change the relationship between OEM and its suppliers for a more collaborative approach, as opposed to a strictly contract-based approach \cite{Knauss2016}. This enables both partners to have tools to better support Continuous Deployment.

\subsubsection{Architecture}
In this section, we will cover the solutions that are related to the architecture of the application. As Shahin et al. found, ``it has been reported by both architecture practitioners and non-architecture practitioners that software architecture should be taken into account, and should be changed and optimized during the journey to adopting Continuous Deployment practice.'' \cite{Shahin2016}. Some papers present principles that can simplify the adoption of Continuous Deployment.

One of the aspect related to the architecture of the application that can help when implementing continuous deployment is having a microservice architecture. This is one of the ways to achieve the principle of ``small and independent deployment units'' \cite{Shahin2016}. We read in many papers that it is harder to migrate a monolithic application than an application with smaller sub modules. The main advantage of having a microservice architecture is that it is a bit easier to test and, when publishing new changes, developers are more confident in what might be impacted. With monolithic architecture, one change can impact a lot of different aspect of the application \cite{Zhu2016}. Also, this architectural style enables the deployment of those microservices independently of others \cite{Shahin2016}. However, in a lot of cases refactoring a monolithic application into a microservice approach can be near impossible \cite{Chen2017}. This is why this solution is targeted towards newer projects that are trying to use continuous deployment \cite{Zhu2016}. Another problem with this solution is that not only the architecture needs to change, but the team structure needs to change to adapt to this paradigm. Some team maturity is needed for this \cite{Shahin2016}.

In order to improve and facilitate the implementation of Continuous Deployment, making sure all microservices are backwards compatible can prove to be very valuable \cite{Callanan2016}. This will help reduce dependency between microservices. It will allow each microservice to be deployable independently without affecting the reliability and stability of the other microservices and the project as a whole. Having backwards compatibility will allow changes and features to be deployed faster if other microservices are not affected by those changes.

Another way to achieve the principle of ``small and independent deployment units'' \cite{Shahin2016} is to consider the database as a continuously deployable unit \cite{Shahin2016}. To achieve that, big databases should be split into smaller databases, but each database should also be deployable as if it was code \cite{Shahin2016}. Those two things can simplify the deployment of a component by removing the database as a bottleneck for an independent deployment.

A somewhat surprising architecture principle that can help implement the Continuous Deployment practice is to not focus too much on re-usability \cite{Shahin2016}. The reason to that is that it increases the dependency between teams, as a reusable component needs to meet the expectations of multiple teams. The coordination between teams is difficult and the team responsible for the ``reusable module'' cannot deploy its module autonomously as it can impact the deployment of other teams \cite{Shahin2016}. Another reason is that it reduces the testability of the component that uses the ``reusable module'' \cite{Shahin2016}.

Another solution regarding architecture is to add logging to the application and then aggregate the log files produced \cite{Shahin2016}. The rationale behind this solution is to be able to monitor the application and also have a way to help detect where are the problems, because the application is being deployed way more frequently with CD. The stability of the application can then more easily be assured. As the continuous deployment tools does not provide a sufficient logging, external tools and logging inside the application itself should be used. Log files can then be aggregated and made search-able. When logging and log aggregation works well, it has been reported that it can help find performance problems and even eliminate the need of rollback.

It is also possible to architect applications to be able to isolate changes. The goal is to ``isolate changes and minimize the impact of changes'' and so ``software units (e.g., component or service) had to be small enough to be modified, replaced and run independently in productions environments'' \cite{Shahin2016}. Teams can then autonomously and independently modify those software units and put them in production. Some ways to achieve that goal is to use \em{domain driven design} and \em{bounded context}, as they help identify components at a business level instead of separating them by functionality. The components of an application based on the business are generally more independent and thus they enable isolated changes. It is also possible to use ``feature toggles`.

One last solution is to build the architecture with testability in mind \cite{Shahin2016}. Keeping in mind to add testability inside the architecture helps to choose tools and technologies that support that. Dependency injection is another way to add testability inside the application. The architecture of the application should also support test automation.

\subsubsection{Process}
This section will go over the solutions related to the development process.

The first solution associated with this category is creating a dedicated team that is going to only focus on implementing the continuous deployment pipeline and process. The idea is that if you assign the implementation of the continuous deployment pipeline to a normal development team, they will only focus on that task part time because they still need to work on the development of their application \cite{Chen2017}. The slow implementation of the deployment pipeline can lead to the project being shut down, so it really is best to focus exclusively on that task \cite{Chen2017}. Another benefit of having a dedicated team to implement Continuous Deployment is that you can create a team that is multidisciplinary, meaning people who work in different sections of a more traditional development process \cite{Chen2017}. Creating the team of developers, but also people from the quality assurance team and people with a security background, helps create a balanced pipeline and can help spotting flaws in the new process \cite{Chen2017}.

The second solution is to apply some continuous practices principles during the implementation of the continuous deployment pipeline. The idea behind this solution is to try to implement the pipeline in small steps, because you do not want to spend a lot of time without any signs of progress \cite{Chen2017}. A Continuous Deployment pipeline contains a lot of different section (e.g. build system, running automated test suite, deploying, etc.), this mean that it can easily be separated into small tasks that are easier to implement \cite{Zhu2016}. For example, you can start by having the pipeline only running a build on the code base, this is easy to implement and it is the first step into creating a functional pipeline \cite{Chen2017}. There are many advantages to the method of implementation. The first advantage is that by implementing small parts of the pipeline at a time, it makes it easier for the teams to adopt it, and progressively start using the pipeline \cite{Chen2017}. The second advantage of the incremental implementation of the deployment pipeline is that you can gather feedback from the development teams quickly, this is at the core of the continuous practices (and Agile process), by release small feature quickly it is easier to adopt the pipeline and to give continuous feedback in order to create the pipeline that is most fitted to the company \cite{Chen2017}. The third advantage is that big companies have broad technology stack (many different programming languages / frameworks) and it makes it easier to start focusing on a specific technology stack rather than all the technologies at the same time \cite{Chen2017}. By focusing on a small portion of the technology stack you can better evaluate if the pipeline works correctly before making it more generic to support different technologies, without losing time \cite{Chen2017}.

The third solution is to first start implementing the continuous deployment pipeline for an application that is going to be relatively easy to migrate, but has a big impact on the company. It is very useful to start on an application that is going to be easy to migrate. That way the team can see the benefits of having a Continuous Delivery pipeline without too much work. You should be looking for an application with a relatively small code base, a more modern architecture, and, if possible, an application that already has a test suite \cite{Chen2017}. Since it is going to be the first application to have a Continuous Delivery pipeline it is not recommended to tackle the biggest application of the company. The second aspect of this solution, is that you want an application that has an impact on the business \cite{Chen2017}. That way, the stakeholders will notice it more and be more incline to continue the switch towards Continuous Delivery.

The fourth solution in this category is deploying features that are not user facing. One challenging aspect of continuous deployment is that every change in the code base is deployed to the customer, this can cause problems when implementing large features in multiple small changes because you do not want the customer to have an half implemented feature. The idea behind this solution is that you can deploy changes whenever as long as it is not user facing. For example, if a big website wants to create a new feature, it can first start by deploying the back end, and then, once the back end is ready, deploy the front end that is going to let the customers use the new feature \cite{Parnin2017}. An unusual side effect of this solution is that the developers will start pushing dead code (code that is never called) while waiting on the front end component to be merged \cite{Parnin2017}.

The fifth solution here is creating a separated process or team to oversight the new features that could have security issues. One of the challenges of the fast paced release, can be the lack of confidence that a feature of secure, when dealing with sensitive user information you need to be very careful not to cause any breach, and knowing if there is a security issue can be challenging for a normal developer. The solution to this problem is to create a team that is dedicated to review patches that could have security holes in them. By doing that, the developer can feel confident and do not waste time trying to analyze their patch on a security level \cite{Parnin2017}. Plus, the patches that could be harmful are analyzed by a team of experts, who are more likely to catch a problem \cite{Parnin2017}. Overall, this helps the momentum of the team and reduced the risks of the company \cite{Parnin2017}.

The sixth solution regarding the process is to use model-driven development when appropriate. This solution enables the organization to automatically test and deploy their models, as in the automotive industry \cite{Knauss2016}.

\subsubsection{Tools}
This section will focus on the solutions that are related to the tools of the continuous deployment pipeline implementation.

The first solution in the category is having a visual representation of the pipeline and the pipeline implementation progress. Migrating large applications can be very challenging and time consuming, this often translate into a lack of motivation and a slow down of the team momentum \cite{Chen2017}. One of the solution to this issue is to have visual representation of what the pipeline will look like, and the current state of the pipeline implementation \cite{Chen2017}. There are two aspect to this solution, the first one is that having a visual representation of what the pipeline should look like helps the team knowing what are the steps that should be taken to ensure the proper implementation of the pipeline \cite{Chen2017}. The second aspect of this solution is that by having a visual representation it is easier to tell which stages of the pipeline are implemented and where is the team standing at. Even if the pipeline is not completely implemented, knowing the team progress can help maintain the team's continuous deployment migration momentum \cite{Chen2017}.

The second solution in the tools category is having the configuration of the pipeline as code. This solution is more of a tip when choosing a continuous deployment tool or when implementing an in-house tool. Having the configuration of every stage of the pipeline as code (most of the time in a format like JSON, YAML or XML) has many advantages over a more traditional UI configuration method \cite{Parnin2017}. The first advantage is that you can put the pipeline configuration in a version control system (like Git or Subversion), this mean that the historic of the configuration is always saved and it is easy to roll back to a previous configuration \cite{Parnin2017}. The second advantage is that it can be peer reviewed easily, since it is just a normal text file it can be easily integrated to a code review platform to make sure that every change in the pipeline is verified \cite{Parnin2017}. The third advantage is that it is a lot easier to find bugs or problems in a configuration file rather than with a graphical user interface; it is easy to miss a configuration option on a graphical user interface \cite{Parnin2017}. The fourth advantage is that it is easier to replicate, if a team need to create a new pipeline for a new application and want to use the same configuration as another pipeline, that is really easy with files, but it is much harder if you have to manually enter the configuration on a user interface \cite{Parnin2017}. To sum up, having the configuration of a pipeline as a text file helps a lot with readability and usability.

The third solution is to use a turn-key solution \cite{Klepper2015}. When a solution without much configuration is available, it has been reported that the usage of a deployment pipeline has increased.

\section{Discussion}
\label{sec:5}
In this section, we go over which challenges can be solved by specific solutions and which challenges we weren't able to find a solution to out of the solutions we have found.

\subsection{Solutions to each challenge}
\begin{figure*}[!t]
\centering
\includegraphics[width=7in]{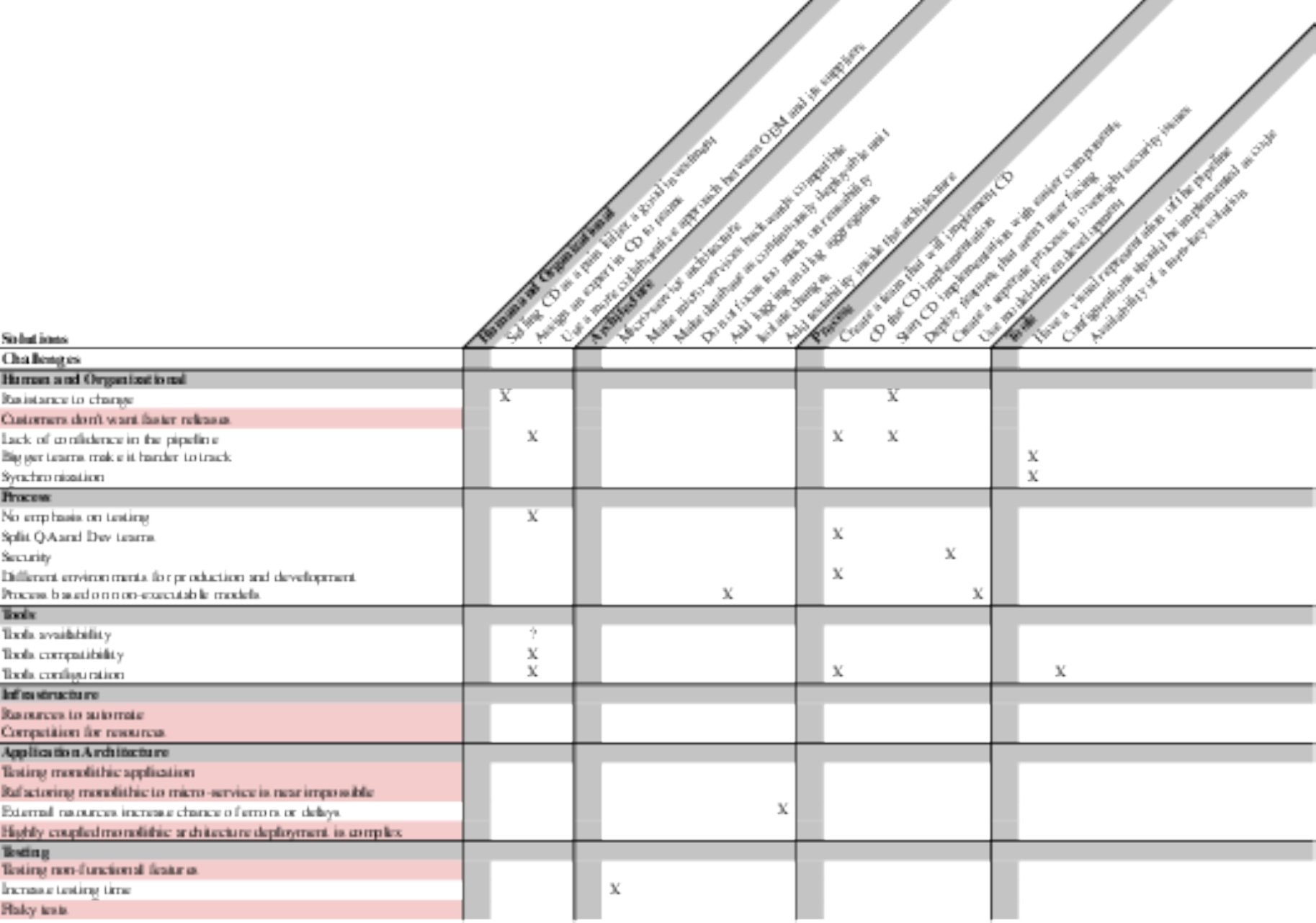}
\caption{Solutions associated to challenges}
\label{figure_challenges_solutions}
\end{figure*}

Out of the 21 challenges we found, 14 of them can be solved by applying some of the solutions discovered
(see Figure \ref{figure_challenges_solutions})
. We will go over each solution and which challenge(s) can be solved by it.

From the results that we have found, we believe that assigning an expert in Continuous Deployment to teams that are implementing Continuous Deployment can help with 5 of the challenges found, such as a lack of confidence in the pipeline, the lack of emphasis on testing, tools availability, tools compatibility and tools configuration. Having an expert in Continuous Deployment directly helping out a team will surely increase the confidence in the Continuous Deployment pipeline as the pipeline will have to be approved by this expert, as this will not be the first time he implements CD. The expert will be able to recommend the right tools to support Continuous Deployment and he will be able to guide the team towards a successful Continuous Deployment implementation. He will not, however, be able to guarantee the development of the application being developed. An expert in Continuous Deployment will know how important testing is in the Continuous Deployment process. Because of this, he should make sure that the testing of the application and each of its component is adequate and will provide an acceptable level of quality. Also, as mentioned above, the tools that will be used to implement Continuous Deployment will be recommended by the expert in CD. The expert will be able to make the right choice as to which available tools are compatible together under the right configuration and will offer the best solution.

Similarly to assigning an expert in Continuous Deployment to a team, it can be good to create a team that will take care of implementing Continuous Deployment for different results. This team can not only help reduce the lack of confidence in the pipeline and manage Continuous Deployment tools' configuration, but it can also help dealing with split QA and development teams as well as managing different environments for production and development. This Continuous Deployment focused team should take care of most of the work related to the Continuous Deployment pipeline which would alleviate the work of the development team related to this process. Because they will be taking care of the pipeline, they will represent a bridge between the development team and the QA team. Over time, the connection between the development and QA team should be achieved without the help of the Continuous Deployment focused team, but they will make this process easier.

Selling Continuous Deployment as a pain killer and a good investment can help convince the team to adopt Continuous Deployment in the presence of resistance to change. A very common argument used against implementing Continuous Deployment is the initial cost and time required for this process. Convincing reluctant employees that Continuous Deployment is a good investment for the future will help reduce the resistance to this change.

The use of microservices for the application architecture can greatly help reducing testing time. Time is always a very important factor in software development and we should always aim to reduce testing time as it reduces the time for a developer to know about an issue in his latest work and therefore, reducing development time. The way a microservice architecture can help with reducing time is simply due to the fact that microservices are more easily testable in a more independent fashion. If a developer is working on a single microservice, there won't be a need to test each and every microservice on its own due to their independence.

Adding logging and log aggregation can help with the challenge that occurs when a process is not based on executable models. Because it is harder to test these non-executable models, the addition of logging will be valuable as it will help monitor the execution of the application once deployed.

Adding testability inside the architecture is another great step to take, especially when dealing with multiple external resources, as they greatly increase the chance of errors or delays in the pipeline. Testability inside the architecture can help deal with external resources as it can help prevent potential errors from those external resources such as unavailability during the testing phase.

When facing resistance to change or a lack of confidence in the pipeline, we believe a good way to solve this is to start the Continuous Deployment implementation with easier components. This way, the team will acquire experience with Continuous Deployment and its implementation by experimenting with said components. This will make the team more comfortable with CD. This should answer some of the initial concerns that the team had and they will learn to have more trust in the Continuous Deployment process.

When security is an important factor within the application, creating a separate process to oversight security issues is a very good practice. Security is often overlooked in a typical Continuous Deployment process. Adding this additional process or team to the Continuous Deployment pipeline and implementing and adapting it correctly to the application can be very beneficial as security concerns can lead to resistance to change.

When appropriate, depending on the application architecture, it is recommended to use model-driven development, in order to avoid ending up with a process that is not based on executable models. Having executable models makes testing and maintaining the tests much easier which makes the process of Continuous Deployment easier as well. 

Having a visual representation of the Continuous Deployment pipeline is a good practice as it will help a lot with tracking and synchronization. In bigger teams, tracking can become tedious when each team member is working on a different part of the application. Having a visual representation of the pipeline will help keep track of the progress of each developer so everyone can see as well as whether or not his changes are accepted by the pipeline. Synchronization is also an issue when multiple parts of the same feature are being developed. For example, in the case where there is a front-end and a back-end part to a feature, the back-end part must usually be deployed first. With the help of a visual representation, it is easy to determine whether the back-end part has been deployed or not, so the front-end part can then be deployed.

Tools configuration in the Continuous Deployment process is a very important factor. Each tool has its own configuration and it should be easily modified, which is why it is recommended to implement configurations as code. Not only will this make their modification easier, it will also allow the team to track those modifications. Some modifications can sometimes alter the pipeline in ways that create unwanted issues, and it is good to be able to backtrack those changes to resolve the issues.

\subsection{Unsolved Challenges}
Unfortunately, we weren't able recommend a solution to every challenge that we have found. 

Three of those challenges are under the Application Architecture category. Those challenges are related to monolithic architecture inside applications, such as testing a monolithic application, the complex deployment of highly coupled monolithic architecture and the difficult or even impossible refactoring of monolithic architecture into microservices. We can take from this that application architecture can be quite important when it comes to implementing a Continuous Deployment process. We can also conclude that not all applications can be developed with the use of a Continuous Deployment process. A monolithic architecture is usually not something that developers want to achieve, but it is sometimes inevitable in some situations. These situations are not compatible with CD.

The next category that offers challenges that we were not able to give a solution to is Infrastructure. In this category, the challenges were about finding and having the resources available to automate the Continuous Deployment pipeline and competition for these resources inside the team. These challenges are directly linked to the budget available to the team and usually cannot be resolved in other ways. So in a way, the solution to these challenges would be to get more budget for the project which the team is not in control of. When it comes to competition for the resources, the solution would be to increase the amount of resources, but this comes down to the same initial challenge of not having enough budget.

Another unsolved challenge is the testing of non-functional features. Non-functional features can cause issues because, as the name implies, we cannot test their functionality. The Continuous Deployment pipeline can rely on unit tests but that is usually not enough to guarantee the software quality.

Last but not least, customers sometimes don't want faster releases, especially when stability and security are important factors to the application. Using live switches would be a possible solution until the customers are ready for the changes, but live switches can create different issues.

Apart from those unsolved challenges, there are also some solutions that are not completely defined and that could be further explored in future studies.

First, for the architecture principle of ``small and independent deployment units'', the research has not found what is small enough to contribute successfully to Continuous Deployment practice \cite{Shahin2016}.

Secondly, the solution to add logging and the aggregate the log have open-ended questions: what is useful data in logging and how to make the logs more readable to operators, so they can use them \cite{Shahin2016}.

Thirdly, designing an application to support test automation does not solve the challenge of automating user acceptance tests \cite{Shahin2016}. Those tests are nonetheless very demanding in terms of human resources.

Fourthly, even if the availability of a turn-key solution is really helpful when implementing Continuous Deployment, there are not that many solutions available \cite{Klepper2015}, so more research have to be done on this subject. Also, some more research needs to be done regarding the integration of the tools within organizations with complex products \cite{Knauss2016}.

\section{Threats to Validity}
\label{sec:6}

This systematic literature review suffers threats to validity. First, the selection of the articles have been done strictly by using a very specific query with keys terms for continuous deployment and filters. The result has been a manageable list of articles to analyze, but for the SLR to be really ``systematic'', it would have been more appropriate to use a technique such as ``snowballing'' to analyze the articles and look for their references and for the articles that references them. The list of articles would have been more comprehensive and more related to research. In fact, some articles found by the query were not really useful, even if they matches the query.

Secondly, the classification of the challenges and the solutions has been done by the authors without a specific methodology. It was based on the classification found in each article, but a formal methodology could have produced a more meaningful classification. This is also true for the grouping of some problems. A problem found in an article sometimes look similarly to one found in some other article, despite being named differently, so the authors chose to group them, but this was not made with a specific methodology. Some problems can then have been erroneously grouped together, while some others may be duplicated.

Finally, the association between the challenges and the solutions were sometimes found in the literature, but sometimes, the authors feel that a solution from one article could be applied to a problem from an other article. That implies that not all of the associations have been validated by the research and can thus be incorrect.

\section{Conclusion}
\label{sec:7}
This paper is a systematic literature review covering the challenges that can emerge when a team is trying to implement continuous deployment, as well as the solutions that can help counter those challenges. Our goal was to identify which solutions can be applied to which challenges. We analyzed a total of 31 research papers that were published after 2015.

From the analysis of those papers we discovered a total of 22 challenges that can be categorized in the following 6 categories: \textit{Human and Organizational}, \textit{Process}, \textit{Tools}, \textit{Infrastructure}, \textit{Application Architecture}, \textit{Testing}. The two categories with the most challenges are \textit{Human and Organizational} and \textit{Process}.

During this research we were also able to identify 19 solutions grouped in the following 4 themes: \textit{Human and Organizational}, \textit{Architecture}, \textit{Process} and \textit{Tools}. The main category for the solutions is \textit{Architecture}.

We successfully associated solutions to 14 out of the 22 challenges. The categories with the most challenges that could not be associated with a solution are \textit{Infrastrucure}, \textit{Application Architecture} and \textit{Testing}. 

\section*{References}
\bibliographystyle{unsrt}
\bibliography{references.bib}

\end{document}